# "Transforming Crises into Opportunities: From Chaos to Urban Antifragility."

*Joseph Uguet[a], Nicola Tollin[b] & Jordi Morató[a]*

[a]*UNESCO Chair on Sustainability, Universitat Politècnica de Catalunya-BarcelonaTech, Spain*
[b]*Southern Denmark University, Denmark*

**Abstract**

Urban crises—floods, pandemics, economic shocks, and conflicts—function as accelerators of urban change, exposing structural vulnerabilities while creating windows for reinvention. Building on a prior theoretical contribution that identified fifteen principles of **urban antifragility**, this paper tests and operationalizes the framework through an empirical assessment of **26 cities** selected for their post-crisis adaptation trajectories. Using a tailored diagnostic methodology, we benchmark cities' **Stress Response Strategies (SRS)** and then evaluate **Urban Development Trajectories (UDT)** across four weighted dimensions, positioning each case along a fragility–robustness–resilience–antifragility continuum and applying a balanced-threshold rule to confirm antifragile status.

Results show that "**resilience enhanced by innovation and technology**" is the most effective response typology (**86.9/100**), and that **six cities** meet the antifragile trajectory criteria. By mapping best practices to activated principles and analysing co-activations, the study identifies a robust "hard core" of principles—**Sustainable Resilience (O), Strategic Diversity (F), Proactive Innovation (I), and Active Prevention (N)**—supplemented by operational enablers (e.g., anticipation, mobilization, shock absorption).

The paper concludes by proposing an evidence-based, SDG-aligned operational model that links high-impact principle pairings to measurable indicators, offering a practical roadmap for cities seeking to convert crises into sustained transformation.

**Introduction**

Crisis are not temporary anomalies in the urban record: they are one of its major protagonists. Floods, pandemics, conflicts, economic collapses, and other disruptions - each upheaval reveals the structural limits of governance models, exposes hidden fragilities, and tests the ability of cities to survive. But history also proves that cities that transcend their crises draw new strength from them, adapt, reinvent themselves, and metamorphose in the long term.

Restoring a previous state is no longer enough when the challenges evolve faster than institutional capacities. It becomes imperative to go beyond passive adaptation and embrace an approach that harnesses uncertainty as a driver of progress, converting it into a source of transformation, so cities can thrive in the face of chaos.

Antifragility (Taleb, 2012) reframes resilience by emphasizing learning through experimentation, adaptive flexibility, and iterative improvement. Several cities under severe stress have shown how constraints can become strategic levers. Rotterdam has turned recurring floods into a platform for water innovation, Tokyo has embedded advanced technologies to bolster resilience to natural hazards, and New Orleans transformed post-Katrina recovery into a model of urban regeneration. Together, these cases illustrate how uncertainty, when deliberately engaged, can produce systems that perform better because of shocks—not despite them.

Building on our earlier theoretical contribution— which proposed a conceptual framework organized around fifteen principles of urban antifragility—this article undertakes an empirical assessment across a sample of 26 cities. We analyse the strategies associated with post-crisis transformation, identify the principles operative in situ, and relate them to observed trajectories of urban outperformance, with the aim of elucidating the mechanisms by which crises are converted into opportunities for sustainable, long-term evolution.

This analysis moves beyond theory to propose an operational framework grounded in lessons from demonstrably successful urban transformations. In this view, antifragility is not an abstract construct but as a strategic lever—a practical architecture for reorienting sustainable urban planning amid the growing uncertainties of the twenty-first century.

**Theoretical principles of urban antifragility: a renewed framework?**

*The role of crises as triggers for urban transformation*

Urban crises serve as stress tests that reveal underlying structural vulnerabilities. Some cities fail in the face of shocks, remaining prisoners of their structural vulnerabilities, while others transform crises into levers of restructuring and systemic change (Davoudi, 2012; Taleb, 2012; dos Passos et al., 2018).

Robust cities absorb shocks without adapting, making them vulnerable when challenges become more dynamic (de Bruijn, 2019); by contrast, **resilient cities** prioritize restoring systems to their previous state, thus maintaining structural vulnerabilities and slowing down any adaptive transformation (Davoudi, 2012).

Despite broad agreement that cities must strengthen their capacity to navigate crises, critiques of resilience converge on several points: conceptual ambiguity (Davoudi, 2012), poor fit with systemic, cascading crises (Munoz et al., 2022), lack of truly inclusive approaches (dos Passos et al., 2018), institutional inertia (Roggema, 2021), reluctance to embrace transformation as an adaptive lever (Taleb, 2012) and the marginalization of locally driven strategies (Wahba, 2021). The imperative is no longer to preserve, but to transform. Urban antifragility offers a dynamic alternative—strengthening a city's capacity to learn, innovate, and restructure through crises (Taleb, 2012). To shed light on how urban systems evolve under pressure, Figure 1 presents four conceptual trajectories - fragile, robust, resilient, and antifragile -, that reflect differing capacities to absorb and respond to shocks. These paths show how cities diverge over time under repeated disturbances. The model underscores how underlying vulnerabilities and strategic responses shape recovery patterns as well as the potential for structural transformation and post-crisis sustainability.

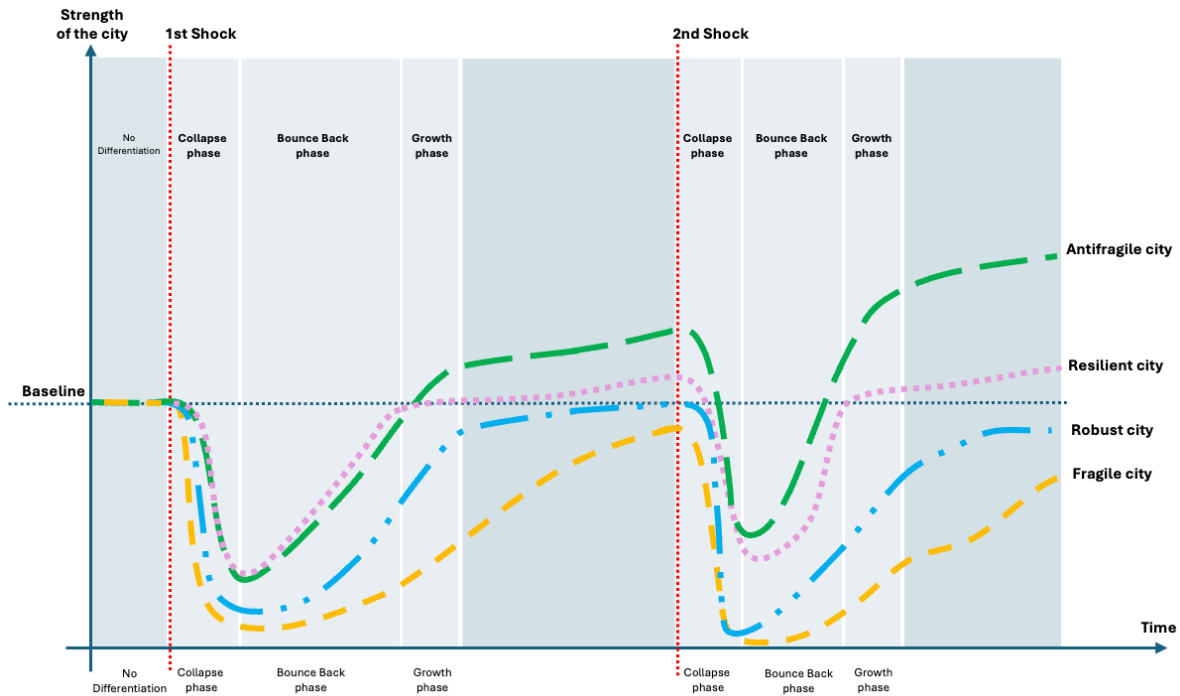

*Figure 1. Urban Trajectories in the Face of Crises: From Fragility to Antifragility. Source: Author*

***Beyond resilience: Synthesizing Urban Antifragility Principles for a Renewed Framework***

In our previous article (*Uguet, J., Morató, J., & Tollin, N., 2025, Post-crisis Strategies: Antifragility Principles as Catalysts for Urban Evolution Towards Sustainability*, *arXiv preprint*), we identified fifteen (15) foundational theoretical principles that structure urban antifragility. This "toolbox" assembles the Antifragile principles that elucidate how urban systems leverage crises as catalysts for improvement. For each principle, it offers a structured summary comprising its code, a concise definition, and a concrete example drawn from the empirical cases.

The proposed framework reveals the critical limitations of urban resilience, which—though essential—often confines cities within rigid, preservation-oriented responses, thus impeding the fundamental transformations required to face increasingly frequent extreme events. In contrast, antifragility prioritizes rapid adaptability and continuous evolutionary improvement, replacing institutional inertia with Systemic Flexibility (B) and Proactive Innovation (I). Rather than merely absorbing shocks, it leverages crises for deep restructuring through Reinforcement through Chaos (H) and Proactive Experimentation (A).

Whereas resilience remains mainly "static" and compartmentalized, antifragility promotes Strategic Diversity (F) and Systemic Anticipation (J), enabling cities to anticipate and adaptively respond to disruptions. It advocates inclusive and resilient governance through Collective Participation (K) and Agile Mobilization (L), while enhancing functional autonomy via Risk Decoupling (G) and Modularity (D).

The main aim of this paper is to test and operationalize an antifragility framework for cities. This conceptual shift reframes urban systems as capable of performing better under stress. Yet, translating theory into practice requires assessing how these principles manifest on the ground. By examining tangible dynamics across a representative panel of 26 cities, we identify the critical drivers of antifragile performance—separating effective levers from secondary ones—and refine the framework for future operational use.

**Methodology**

We apply a tailored, indicator-based diagnostic framework to assess cities' crisis-response performance and position them along a fragility–antifragility continuum. By linking high-performing practices to paired theoretical principles, the method reveals the strategic configurations and levers that most consistently drive urban transformation under stress. Figure 2 summarizes the methodological process designed to reveal, compare, and validate the antifragile dynamics of cities through a four-step analytical framework that articulates (i) empirical benchmarking, (ii) Stress-Response-Strategy assessment (SRS), (iii) Urban Trajectory Diagnosis (UDT), and (iv) validation of the key transformation drivers.

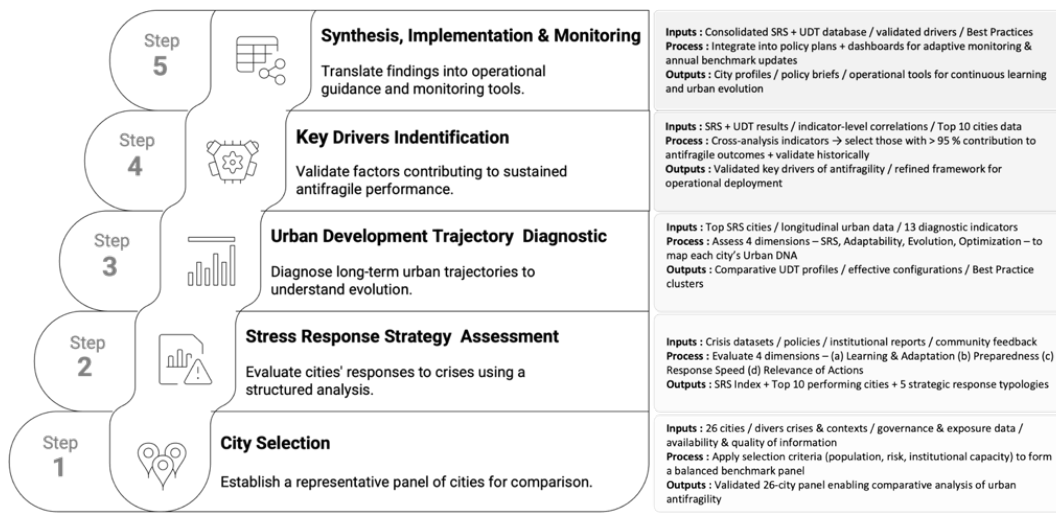

*Figure 2: Five step methodological process. Source: Author*

## City selection: A diverse panel for benchmarking Stress Response Strategies (SRS)

The study draws on a **panel of 26 representative cities selected for their exposure to major crises**, the breadth of their vulnerabilities (climatic, economic, and social), and a balanced distribution across geographic regions and economic profiles. It aims to examine their Stress Response Strategies (SRS) and to identify the factors and pathways that have accelerated—or constrained—their transformation over time. Figure 3 presents the selected case studies and illustrates the diversity of contexts covered in the analysis.

The assessment of Stress Response Strategies (SRS) applies a tailored analytical framework developed to measure and identify the most effective urban responses to crises. The resulting **SRS index consolidates 12 indicators grouped into four dimensions**. Each indicator is scored on a 0-10 scale, reflecting both the degree of implementation (0–5) and the level of impact (6–10) achieved by each measure. The first range (0–5) assesses the operational maturity (from intent and preparation to systemic deployment) while the upper range (6–10) captures the transformation capacity (from operationalization and institutionalization to antifragile regeneration). This dual progression allows distinguishing initiatives that merely exist from those generating measurable, regenerative effects on the urban system. Scores are then aggregated and normalized to produce a composite value out of 100. This overall score captures a city's capacity to respond, adapt, and transform under crisis conditions.

Table 1 presents the **SRS framework, structured around four analytical dimensions**. First, *learning and adaptive capacity* assesses how well cities translate lessons from past crises into concrete policy, governance and infrastructural improvements. Second, *crisis preparedness level* evaluates anticipatory capacity, including planning, resource availability, and the effectiveness of operational protocols. Third, the *speed of response* measures how rapidly urban systems mobilize resources and coordinate across institutions when shocks occur.

Finally, *relevance of actions taken* examines the appropriateness and efficiency of implemented measures -i.e., whether strategies address territory-specific needs and contribute to sustainable, long-term transformation.

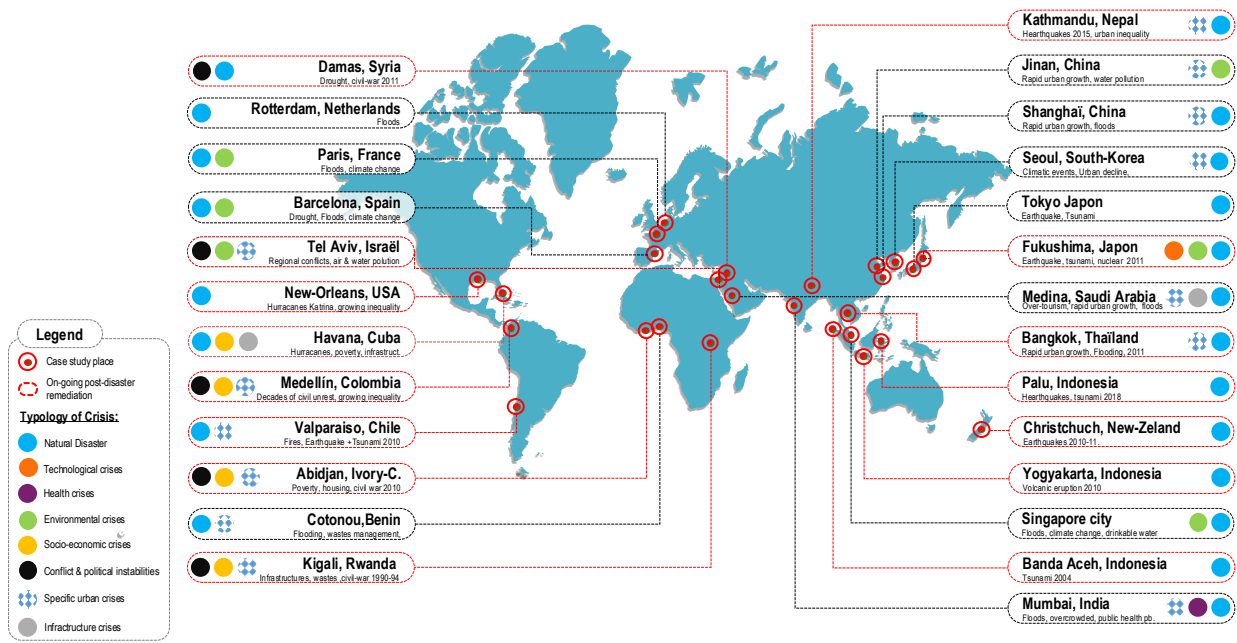

*Figure 3. Panel 26 cities: visualization of their main characteristics. Source: Author*

| Dimension | SRS Evaluation parameters | Description |
| --- | --- | --- |
| Ability to learn & adapt | - Responsiveness to previous teaching<br>- Updating emergency plans<br>- Innovation in responses | Assesses how cities learn from past crises and adapt their strategies to better cope with future crises. Measures resilience and the ability to evolve in crisis contexts. |
| Level of Crisis Preparedness | - Availability of contingency plans<br>- Training and public awareness<br>- Infrastructure and resources available | Assesses how cities prepare for potential crises by putting in place contingency plans, training and raising awareness, and ensuring the availability of necessary infrastructure and resources. |
| Rapid response to crises | - Initial response time<br>- Resource mobilization<br>- Communication and initial coordination | Assesses how quickly cities respond to crises as soon as they arise. Measures the initial effectiveness of responses and the ability to minimize the immediate negative impacts of crises. |
| Relevance of the actions taken | - Alignment with local needs<br>- Effectiveness of interventions<br>- Flexibility and adaptability of responses | Assesses the extent to which the actions taken by cities to respond to crises are appropriate and effective in relation to the specific challenges they face. Analyses the quality of decisions taken and their implementation. |

Table 1. Analytical framework for:Stress Response Strategies (SRS). Source: Author

Figure 4 presents these strategies, grouped into five response typologies: (1) **Resilience enhanced by innovation and technology** (advanced technological integration); (2) **Community participation and governance** (citizen engagement and inclusive governance); (3) **Adaptation to natural and environmental disasters** (post-crisis preparedness and rebuilding); (4) **Sustainable Urban Development & Resilience** (long-term sustainability-aligned strategies); and (5) **Socio-economic resilience and post-crisis reconstruction** (economic revitalization and social support).

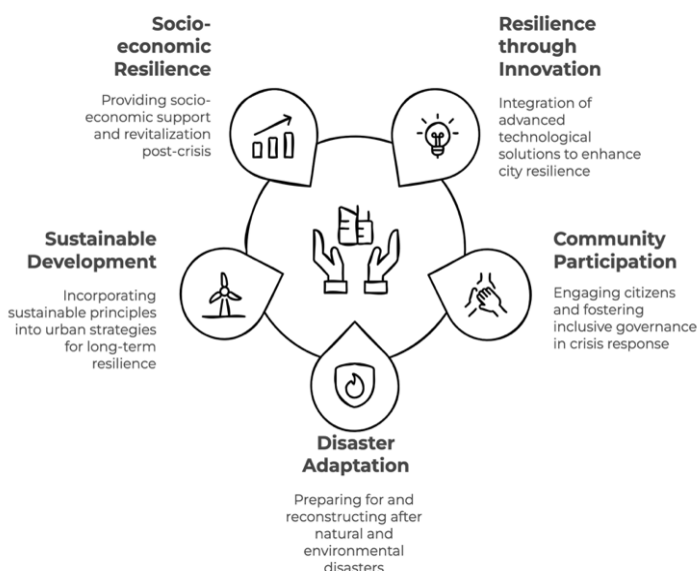

*Figure 4. Five response typologies. Source: Author*

*Diagnostic of the Urban Development Trajectories (UDT)*

At the conclusion of this first phase, the analysis focuses on the ten highest-performing cities, selected based on their overall scores. The evaluation uses a proprietary, heuristic diagnostic tool designed to model urban development trajectories in a simplified yet coherent way, positioning each city along a continuum from fragility to antifragility. The framework combines 13 indicators across four weighted dimensions: *Stress Response* (100 pts), *Adaptability* (60 pts), *Evolution* (60 pts), *Optimization* (60 pts). The total score (280 pts) is converted into a UDT index out of 100, placing cities along the Fragile (<44) → Robust (45-64) → Resilient (65-84) → Antifragile (>85) continuum. To qualify as antifragile, each dimension must exceed 85 % of its maximum score, ensuring balanced systemic performance rather than excellence in only one area.

The **Stress Response Strategy (SRS)** assesses Responsiveness, Preparedness, and the contextual Relevance of interventions during crises. **Adaptability** measures a city's capacity to adjust policies, foster innovation, and sustain continuous learning. **Evolution** assesses the extent of post-crisis structural transformation and long-term system improvement. **Optimization** focuses on resource efficiency and the reduction of systemic vulnerabilities.

As shown in Table 2, these dimensions are operationalized through **thirteen targeted indicators**. The analysis begins with a panoramic review of dominant combinations across the full panel, then shifts to a finer-grained examination of the levers driving indicator outperformance. This second step identifies the strategic configurations most consistently associated with higher scores and, ultimately, the most effective in activating antifragile dynamics.

| Dimensions | Indicators | Explanation |
| --- | --- | --- |
| **Stress Response Strategy (SRS)** | - Ability to learn and adapt<br>- Level of preparedness<br>- Speed of response to crises<br>- Relevance of actions taken | Assesses the effectiveness of strategies implemented by cities to respond to crises. These parameters measure the responsiveness (flexibility) of cities, their state of readiness, their ability to react quickly and the adequacy of decisions in a context of major stress. |
| **Adaptability** | - Ability to adjust policies<br>- Continuous innovation<br>- Strategies for self-organization & self-learning | Examines a city's ability to reorient itself (policies, governance), & adapt their strategies to new challenges to innovate, and encourage collective dynamics including the encouragement of self-organization and continuous learning. |
| **Evolution** | - Post-crisis transformation capacity<br>- Continuous improvement of infrastructure & systems<br>- Longitudinal sustainable development | Analyses the ability of cities to evolve after a crisis, to transform & improve their structures and systems in a sustainable way over time, offering a longitudinal view of efforts. |
| **Optimization** | - Effective use of human, economic & financial resources<br>- Natural resource management<br>- Vulnerability reduction strategy | Assesses how the city optimizes the use of its resources (human, financial, natural) to maximize resilience, while seeking to minimize vulnerabilities. |

*Table 2. The diagnostic tool for Diagnostic of the Urban Development Trajectories (UDT)*. Source: Author

*Identification of key drivers and validation of antifragile dynamics*

During this phase, the analysis identifies the key drivers of antifragile cities, along with the best practices and underlying principles associated with high performance. It characterizes the "**urban DNA" of each city** by tracing how specific strategies translate into measurable outcomes. Each best practice is linked to a binomial theoretical construct—pairing a **dominant** principle with a **supporting** principle—to produce a structured cartography of interactions. This layered mapping clarifies the mechanisms behind strategic results by making visible the relationships between activated principles and the initiatives implemented.

This final analysis focuses on cities with a **validated antifragile trajectory, examining only those indicators that reach at least 95% of their maximum score** in order to isolate the most decisive factors. This stringent filter ensures that the dynamics observed are rooted in empirically exceptional practices rather than average performance. By **mapping each identified best practice to specific antifragile principles**, clear correlations emerge between implemented actions and the adaptive mechanisms they activate.

Restricting the assessment to these top-performing indicators enables a precise identification of the principles that effectively drive antifragility—highlighting how they help convert crises into catalysts for systemic evolution.

The approach also reveals non-intuitive configurations: some principles previously treated as secondary prove decisive, while others assumed to be critical show only marginal influence in practice. By cross-analysing these findings against distinct urban strategy profiles, the truly structuring drivers of antifragility are clarified, thereby refining the framework for future operational deployment.

**Results**

*Effectiveness of Stress Response Strategies (SRS)*

This section synthesizes the evaluation results around two focal points: the **ten strongest-performing cities in crisis response** and the **strategic patterns** underlying their performance. The applied filter brings the most effective approaches into focus and shows how antifragile principles support resilient urban trajectories. Cities are ranked using the SRS Index, allowing the initial panel to be narrowed to those exhibiting the highest adaptive effectiveness. The resulting typologies, once standardized, will provide the basis for a comparative analysis of their relative impacts and for identifying the most effective combinations of strategies.

Figure 5 presents a multi-layered alluvial diagram, mapping the cities according to their economic classification (World Bank, 2024), crisis typology (Figure 3 (legend)), and dominant response strategy (section 3 methodology for details). It highlights correlations between wealth levels, crisis profiles, and the types of strategies deployed. In parallel, Figure 6 offers a global ranking of the SRS scores, allowing a visual comparison of cities' performance and the strategic patterns that emerge from their responses to disruption.

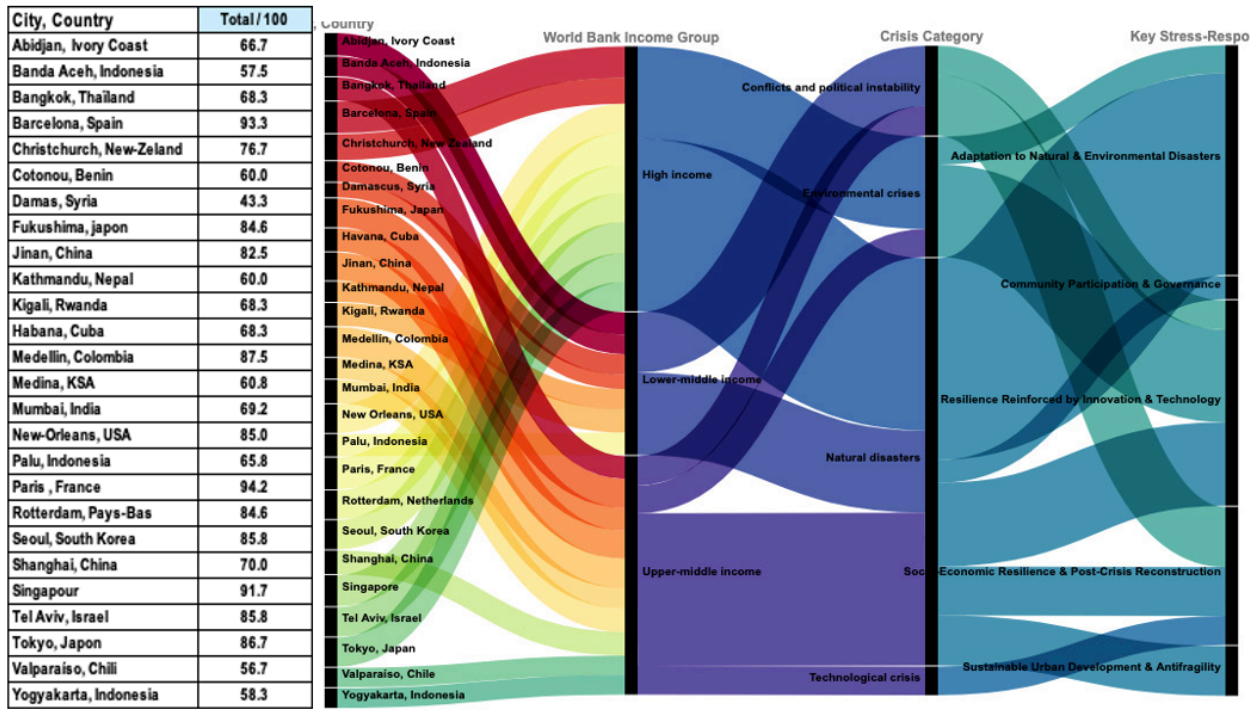

*Figure 5. Ranking SRS Index by country, in "Alluvial Diagram". Source: Author*

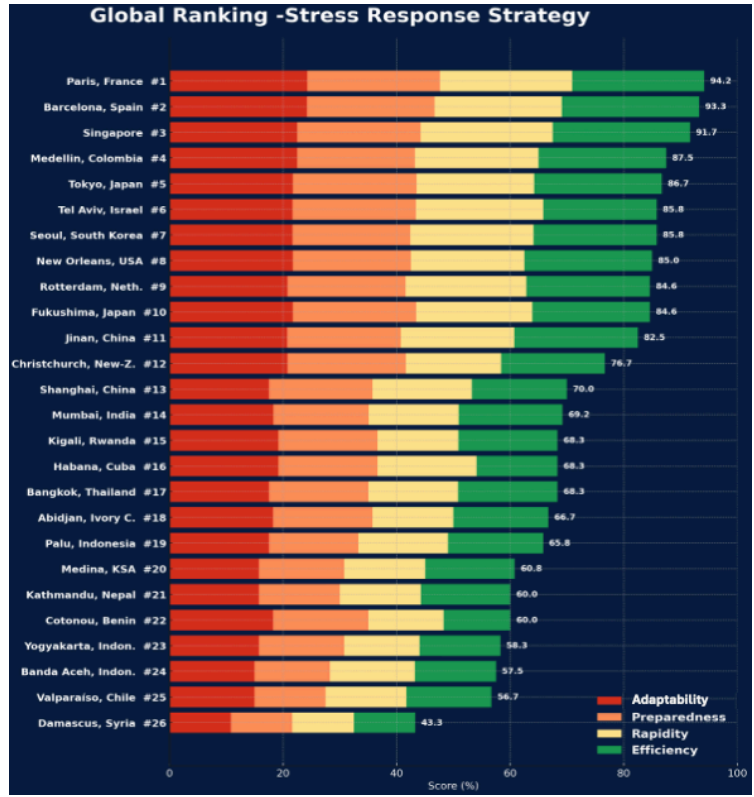

*Figure 6. Global ranking of the Stress Response strategy. Source: Author*

The comparative analysis of the 26 cities reveals substantial differences in strategic effectiveness. Figure 7 shows the distribution of effectiveness scores across the five strategic typologies: the blue boxes represent the interquartile range, the red lines indicate the median, and the yellow whiskers show the minimum and maximum values, with outliers displayed as individual points.

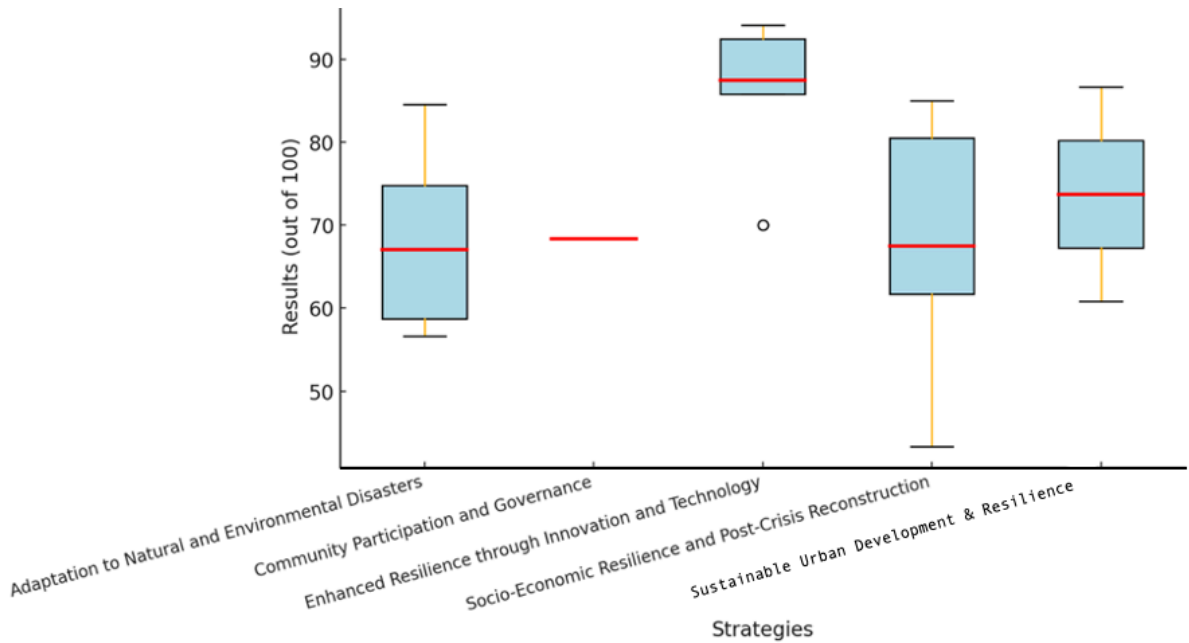

Figure 7. Distribution of effectiveness across strategic typologies . Author: *J.Uguet*

Notably, among the five SRS strategic categories, **'Resilience Enhanced by Innovation and Technology'** shows the highest overall effectiveness (86.9/100) (Figure 8). It is adopted by seven cities, five of which appear in the top ten, corresponding to a 71% top-performance rate. Particularly well suited to environmental and technology-related shocks, this topology is most prevalent in economically strong regions, underscoring the decisive role of innovation in enabling post-crisis transformation.

By contrast, **'Adaptation to natural and environmental disasters'** is the most frequently adopted approach, yet it also registers the lowest average effectiveness (Figure 7 and Figure 8).

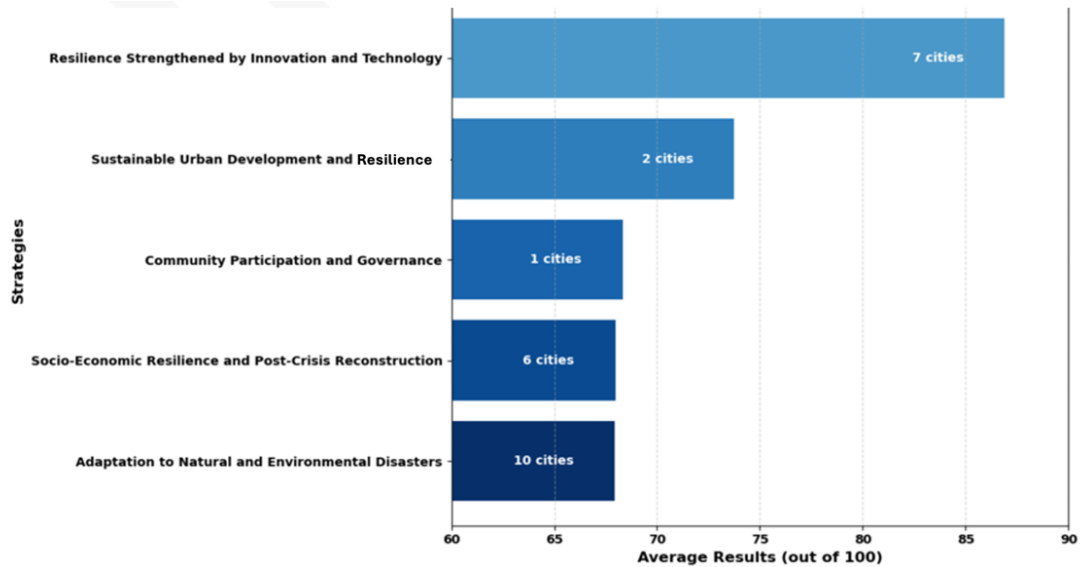

*Figure 8. Effectiveness of stress response strategies. Author: J.Uguet*

Table 3 presents the ten highest-performing cities on the SRS index and the strategies they adopted. Supported by advanced infrastructure, these territories are able to scale technological integration efficiently, strengthening both shock absorption and long-term transformation. Among the top ten, six cities follow an innovation-driven strategy, underscoring its relevance in contexts marked by environmental stress and infrastructural disruption. With the exception of Medellín, all are classified as high-resource (HR) cities, having mobilized substantial means to deploy adaptive technologies. Their experience highlights innovation as a critical lever for turning crises into opportunities for urban development. However, strong strategic effectiveness does not, on its own, guarantee a sustainable trajectory. The next section therefore examines how these trajectories unfold over time.

| City, Country | Stress Response Strategy (Main) | Total / 100 |
|---|---|---|
| **Paris, France** | Resilience Enhanced by Innovation and | 94.2 |
| **Barcelona, Spain** | Resilience Enhanced by Innovation and | 93.3 |
| **Singapore** | Resilience Enhanced by Innovation and | 91.7 |
| **Medellin, Colombia** | Socio-Economic Resilience and Post-Crisis | 87.5 |
| **Tokyo, Japan** | Sustainable Urban Development and | 86.7 |
| **Seoul, South Korea** | Resilience Enhanced by Innovation and | 85.8 |
| **Tel Aviv, Israel** | Resilience Enhanced by Innovation and | 85.8 |
| **New Orleans, USA** | Socio-Economic Resilience and Post-Crisis | 85.0 |
| **Fukushima, Japan** | Socio-Economic Resilience and Post-Crisis | 84.6 |
| **Rotterdam, Neth.** | Adaptation to Natural and Environmental | 84.6 |

*Table 3. Ten highest SRS scores (city ranking). Source: Author*

*Diagnostic of the Urban Development Trajectory (UDT)*

Using our in-house model, we assessed the Urban Development Trajectories (UDT) of each city, positioning them along the fragility–robustness–resilience–antifragility spectrum. This assessment clarifies whether a city's strategies generate genuine antifragile dynamics—or whether they primarily strengthen shock absorption without producing structural change.

Table 4 ranks the top ten cities in the panel, detailing their scores across each dimension of the UDT framework and indicating the corresponding development level. The evaluation identifies six cities with an antifragile trajectory. Importantly, this classification depends on more than a high overall score: each dimension must exceed 85% of its maximum value (Table 4 and Figure 9). New Orleans illustrates this rule, reaching the antifragile threshold despite a lower total score than Medellín. **Antifragility is therefore defined by balanced strength across all dimensions, rather than by aggregate performance alone.**

| Dimensions /Cities | SRS*/100 | Adaptability/60 | Evolution/60 | Optimization/60 | **UDT/100** | Level |
|---|---|---|---|---|---|---|
| **Singapore** | 92 | 56 | 55 | 55 | **92.1** | ANTIFRAGILE |
| **Paris, France** | 94 | 56 | 54 | 51 | **91.1** | ANTIFRAGILE |
| **Barcelona, Spain** | 94 | 55 | 54 | 51 | **90.5** | ANTIFRAGILE |
| **Tokyo** | 87 | 53 | 53 | 54 | **88.1** | ANTIFRAGILE |
| **Fukushima** | 85 | 52 | 55 | 54 | **87.8** | ANTIFRAGILE |
| **Medellín, Colombia** | 88 | 50 | 54 | 50 | **86.3** | RESILIENT |
| **New Orleans** | 85 | 51 | 52 | 51 | **85.4** | ANTIFRAGILE |
| **Tel Aviv, Israel** | 86 | 53 | 50 | 50 | **85.4** | RESILIENT |
| **Rotterdam, Neth.** | 85 | 52 | 53 | 49 | **85.2** | RESILIENT |
| **Seoul, South Korea** | 86 | 53 | 50 | 49 | **85** | RESILIENT |

*Table 4. Ranking of the top 10 cities: UDT scores and urban development classification. Source: Author*

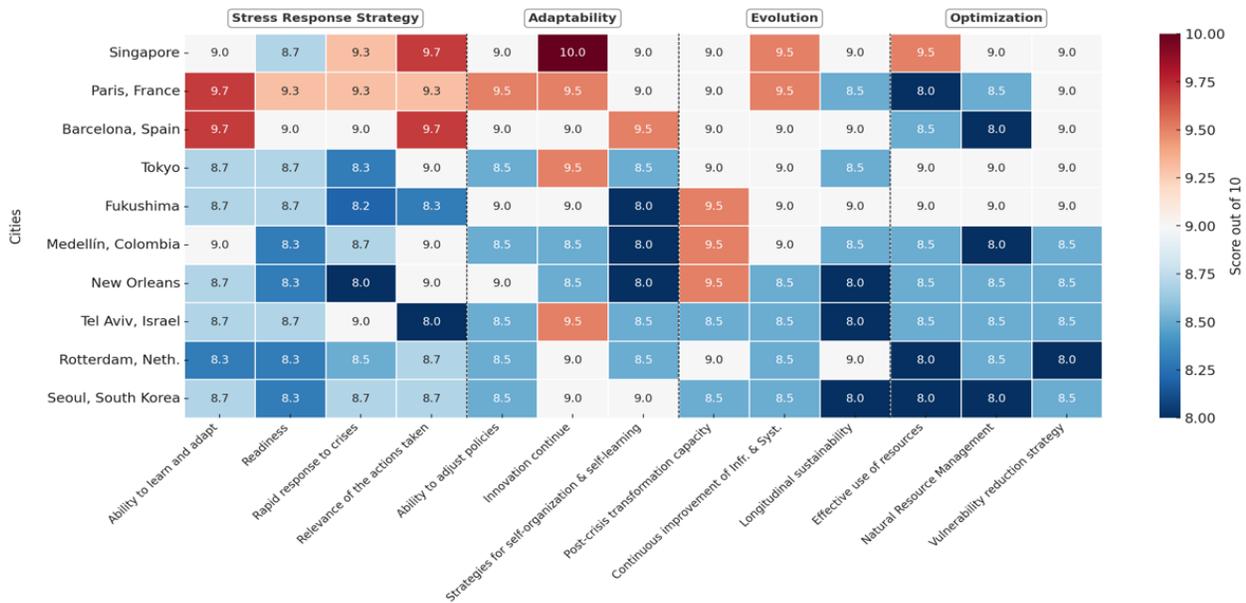

*Figure 9. Heatmap: visualization of the scores by indicator for each city (score out of 10). Source: Author*

Scores are normalized out of 100. Stress Response Strategy holds a higher weight (100/280), while other dimensions were adjusted proportionally (60/100).

*Antifragile Urban Trajectories: Exemplary Cities and the Drivers of Transformation*

**Exemplary Urban Dynamics: Singapore, Paris, and Barcelona Lead the Way.** The evaluation identifies **Singapore, Paris, and Barcelona** as leading examples of coherent transformation toward antifragility (Figure 9 & Table 4). Across the UDT dimensions, these cities combine **innovation**, **learning**, and **infrastructure renewal** as mutually reinforcing structural levers.

**Singapore** stands out as an archetype of antifragile urbanism. Over six decades, it has converted severe constraints—limited natural resources and drinking water, compounded by climate and geopolitical pressures—into development assets through sustained innovation (**10/10**), agile policymaking (**9/10**), and strong resource management (**9.5/10**). **Paris** and **Barcelona** follow closely, with consistently high results in learning (**9.7**), innovation (**9.5**), and infrastructure performance (**9.5** and **9**, respectively). Paris leveraged the momentum of the **2024 Olympic Games** to accelerate a sustainability-oriented transition, while Barcelona—an early mover in institutional resilience since **2013**—has consolidated its role as a testbed for urban sustainability, notably through its participation in **UN-Habitat's CRPT program**.

**Resilient cities facing structural limits: Tel Aviv, Medellín, Seoul, and Rotterdam.** While these cities score strongly in learning and innovation (**8.5–9.5**), their weaker results in optimization (**49–50/60**) reveal the ceilings of resilience-centric strategies (Figure 9). Although effective in crisis response (SRS), they show gaps in deeper systemic transformation and long-term integration. Their resource management scores (**8–8.5**) indicate solid performance yet also point to remaining potential to consolidate sustainable urban evolution.

**Innovation as a catalyst for post-crisis transformation.** The analysis confirms a strong association between innovation and antifragile trajectories. Top performers consistently score **9–10** on Innovation, highlighting the strategic value of technological deployment and experimentation. By contrast, resilient cities often remain comparatively reactive and struggle to translate response capacity into structural change.

**New Orleans** illustrates the importance of balanced systemic alignment. Despite a lower overall score (**85.4**) than **Medellín** (**86.3**), it reaches the antifragility threshold due to stronger coherence across dimensions. This further confirms that **antifragility is not defined by aggregate performance alone, but by sustained strength across all dimensions.**

Overall, the results indicate that antifragile trajectories emerge from the convergence of three interdependent dynamics: **learning**, which anchors adaptation in experience; **innovation**, which turns disruption into creative momentum; and **structural transformation**, which ensures each crisis leaves the city stronger than before. Together, these mechanisms form the architecture of sustainable urban evolution. To move from diagnosis to action, the next section therefore isolates and examines the **best practices** that most effectively activate antifragile principles—laying the groundwork for a scalable, evidence-based framework to guide cities toward **adaptive, replicable,** and **measurable** antifragility.

*Analysis of Best Practices and associated Antifragile Principles (pairs)*

We examined the six cities classified as Antifragile and analysed the strategic practices that shaped their trajectories. In each case, cities implemented a focused portfolio of actions aligned with core antifragility principles, and these alignments proved decisive in explaining high performance.

**Key principles and dominant trends.** Across the six antifragile cities, the analysis shows a consistent activation of principles through the identified best practices (Figure 10). Six principles account for **62%** of all activations. **Principle O (Sustainable Resilience)** ranks first (**20%**), highlighting circularity, efficiency, and system optimization. **F (Strategic Diversity)** and **K (Collective Participation)** follow (**12%** each), reflecting the importance of flexibility and collaborative governance. **I (Proactive Innovation)**, **N (Active Prevention)**, and **M (Social Resilience)** (each **10%**) reinforce anticipation, technological integration, and community capacity. Together, these principles form the structural core of antifragile trajectories by balancing long-term vision with adaptability, inclusion, and innovation.

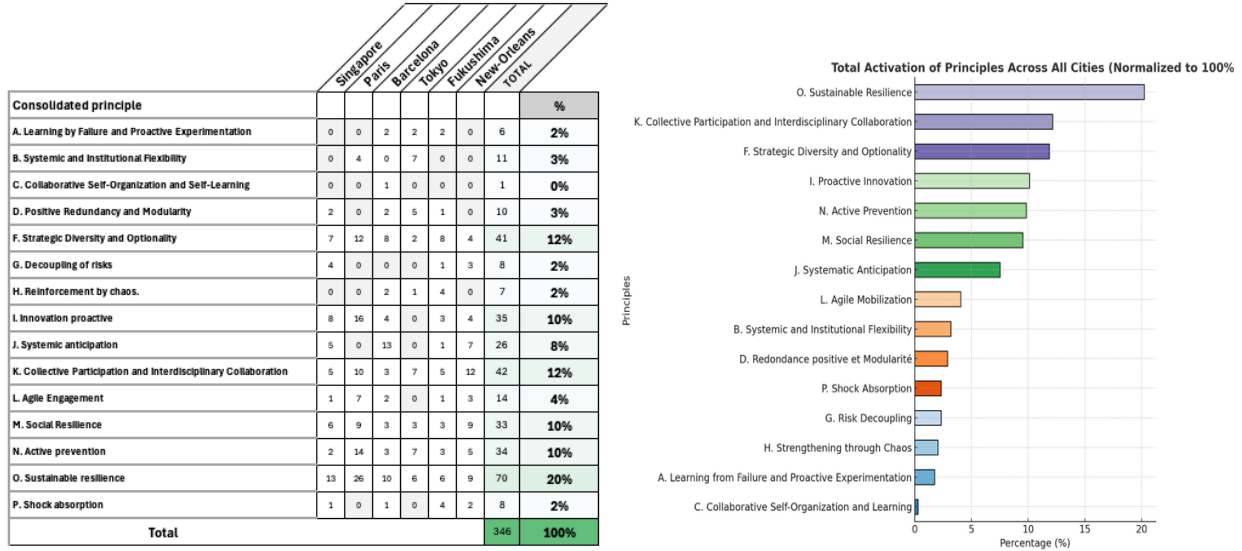

*Figure 10. Best Practices and Principle Activation per Antifragile City. Source: Author*

**City-specific configurations of principles.** Beyond the shared trends, each antifragile city displays a tailored configuration that reflects its context (Figure 10 and Figure 11). **Paris** (main activations: **O=26, I=16, N=14**) combines sustainable resilience with innovation, reinforced by large-scale transformation programs such as the **2024 Olympics**. **Singapore (O=13, I=8, F=7)** articulates a compact "triptych" centered on innovation, modularity, and resilient system design. **Barcelona (J=13, K=10, F=8)** emphasizes anticipation, inclusive governance, and strategic diversity. **Tokyo (B=7, K=7, N=7, O=6)** balances adaptive governance with prevention and collective engagement. **Fukushima (F=8, O=6, K=5)** grounds recovery in participatory diversity and progressive learning. **New Orleans (K=12, M=9, O=9, J=7)** prioritizes community-led reconstruction coupled with longer-term resilience building.

**Adaptive configurations and foundational principles.** Antifragility emerges through context-specific combinations rather than a single replicable model (Figure 11). Across cases, **O, F, K, I, M, and N** form a foundational core. When jointly activated through targeted best practices, these principles enable crises to become structural levers for transformation. These results set the stage for examining **principle pairings** (Figure 12) and the role of **outperforming indicators** in sustaining antifragile dynamics.

**Principle pairings and strategic patterns.** The analysis of principle pairs reveals recurring combinations that shape antifragile strategies (Figure 12). Three dominant duos stand out: **F–O'** (diversity–resilience), **N–O'** (prevention–resilience), and **K–M'** (participation, social resilience). Their repeated occurrence—often also appearing in reverse order—confirms their structuring role in antifragile urban configurations.

**Principle dynamics and strategic weight. Principle O (sustainable resilience)** emerges as a central axis, forming four key pairings and reinforcing its role in circularity and long-term transformation. **Principle I (proactive innovation)** appears in three city-specific pairings, while **J (systemic anticipation)**—also present in three—demonstrates adaptability across diverse contexts. Notably, some low-frequency principles, such as **A (learning from failure)** and **P (shock absorption)**, exert disproportionate influence through specific binomials. Conversely, the absence of **M, L, B, C, G, and H** from dominant pairings raises an interpretive question: are these latent drivers expressed through other mechanisms, or are they theoretically overstated? Linking them to outperforming indicators is necessary to clarify their actual contribution.

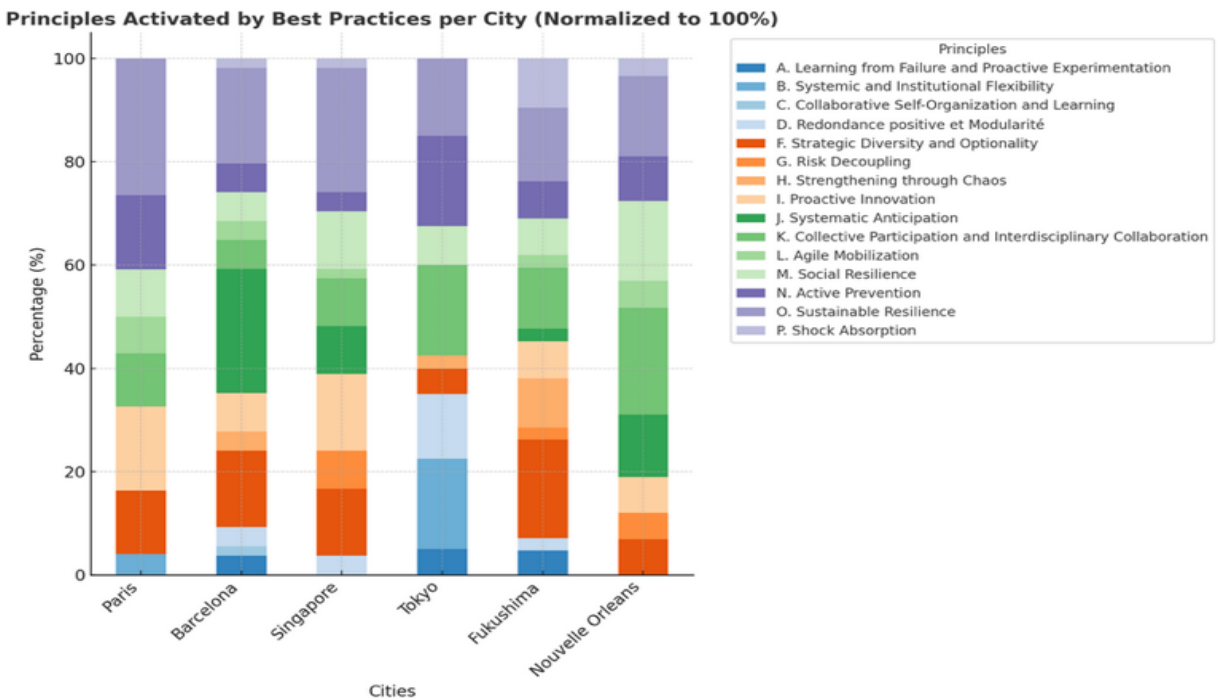

*Figure 11. Distribution by Theoretical Principles of the Best Practices per of each Antifragile City. Source: Author*

| Antifragiles Principals | Paris | Barcelona | Singapore | Tokyo | Fukushima | New Orleans |
|---|---|---|---|---|---|---|
| A. Learning by Failure and Proactive Experimentation | | | | | A-F' | |
| B. Systemic and Institutional Flexibility | | | | | | |
| C. Collaborative Self-Organization and Self-Learning | | | | | | |
| D. Positive Redundancy and Modularity | | | | D-B' | | |
| F. Strategic Diversity and Optionality | F-O' | | F-O' | | | |
| G. Decoupling of risks | | | | | | |
| H. Reinforcement by chaos. | | | | | | |
| I. Innovation proactive | I-F' | | I-O' | | | I-J' |
| J. Systemic anticipation | | | J-F' | J-K' | | |
| K. Collective Participation and Interdisciplinary Collaboration | K-M' | | | | | K-M' |
| L. Agile Engagement | | | | | | |
| M. Social Resilience | | | | | | |
| N. Active prevention | N-O' | | | N-O' | | |
| O. Sustainable resilience | O-N' | O-F' | | | O-F' | O-G' |
| P. Shock absorption | | | | | P-O' | |

*Figure 12. Distribution of the Theoretical Principle pairs per Antifragile city. Source: Author*

### *Principles behind Outperformance: Signals of Antifragility.*

At a finer scale, the analysis isolates indicators that exceed **95%** of their maximum score (Figure 13) to test whether particular principles or pairings consistently drive outperformance. This zoom-in approach shows how principle–practice interactions elevate specific indicators, offering deeper insight into the mechanisms that sustain antifragile urban trajectories.

Across the six antifragile cities, **eight of the thirteen indicators** are activated, with an uneven distribution across dimensions (Figure 13). **Adaptability** is fully mobilized, confirming a shared capacity for adjustment, whereas **Optimization** is less consistently activated. **Continuous Innovation** is the most prevalent indicator, appearing in **50%** of cases and reaffirming its central role in proactive adaptation. Other high-activation indicators—**Learning and Adaptation**, **Post-Crisis Transformation**, and **Relevance of Actions**—also emerge strongly, particularly in **Fukushima** and **New Orleans**. **Singapore** and **Paris** lead the panel, each activating four indicators and showing consistent strength in innovation and system improvement. Figure 14 illustrates the Paris case, showing the principles activated by the indicators that outperform.

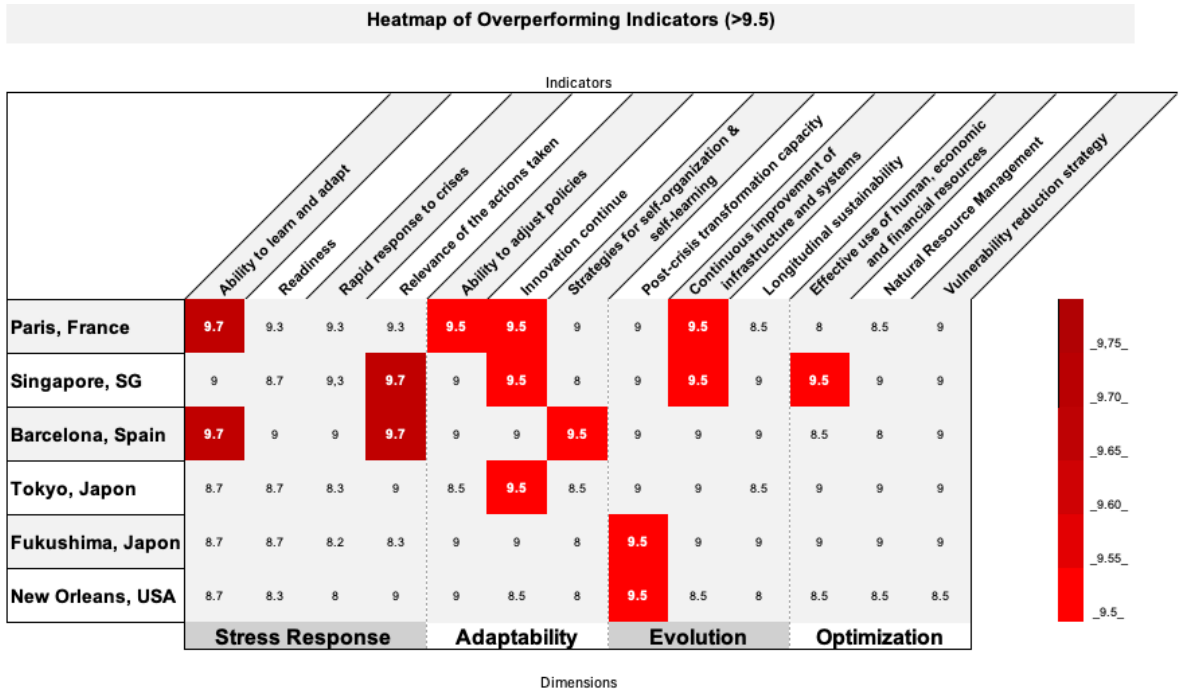

*Figure 13. Mapping of outperforming indicators by antifragile city. Source: Author*

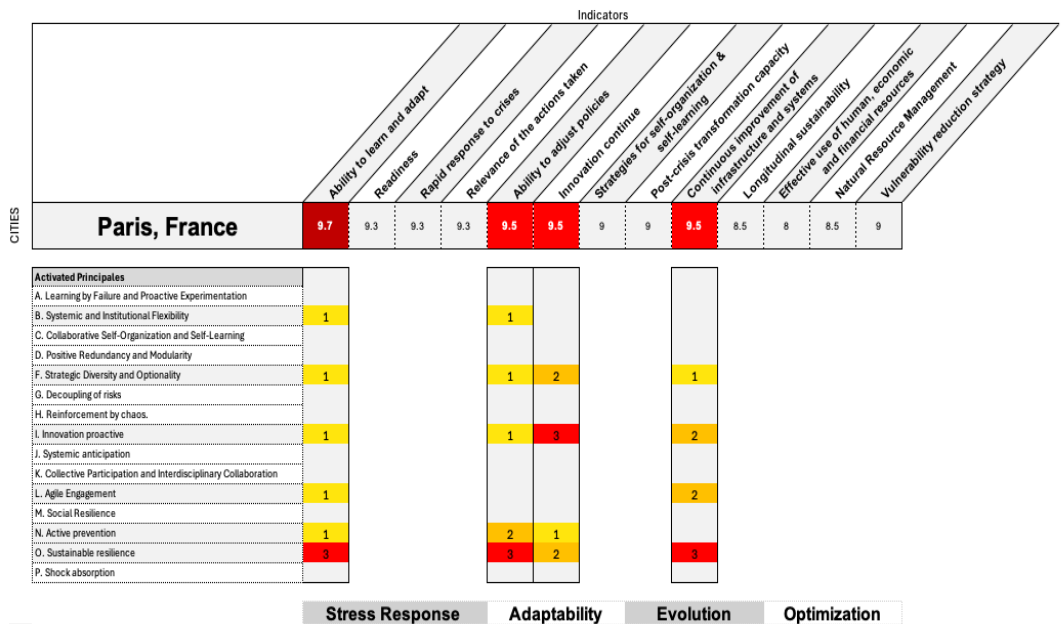

*Figure 14. Paris case study: Principles activated by outperforming indicators Source: Author*

Examining the principles activated by outperforming indicators reveals distinct and strategic contributions to urban antifragility. These high-performing indicators act as a diagnostic lens, isolating the levers most directly associated with superior outcomes across the cities studied.

Overall, the results confirm **innovation, learning, and post-crisis transformation** as core levers of antifragility.

The next analytical step is to identify which Antifragile Principles most consistently activate these indicators, in order to clarify the systemic dynamics underpinning antifragile performance. In this regard, **Continuous Innovation** emerges as a strategic driver. Activated in **Singapore, Paris, and Tokyo**, the Continuous Innovation indicator represents **21%** of all principle activations (Figure 15), highlighting its transversal role in anticipating vulnerabilities and strengthening adaptive capacity. It is driven primarily by **Principle O (Sustainable Resilience)**, followed by **I (Proactive Innovation)** and **F (Strategic Diversity)**. Together, these principles enable rapid adjustment under uncertainty. Notably, **F** accounts for nearly **20%**, underscoring the importance of strategic flexibility for innovation to translate into sustained performance (Figure 16).

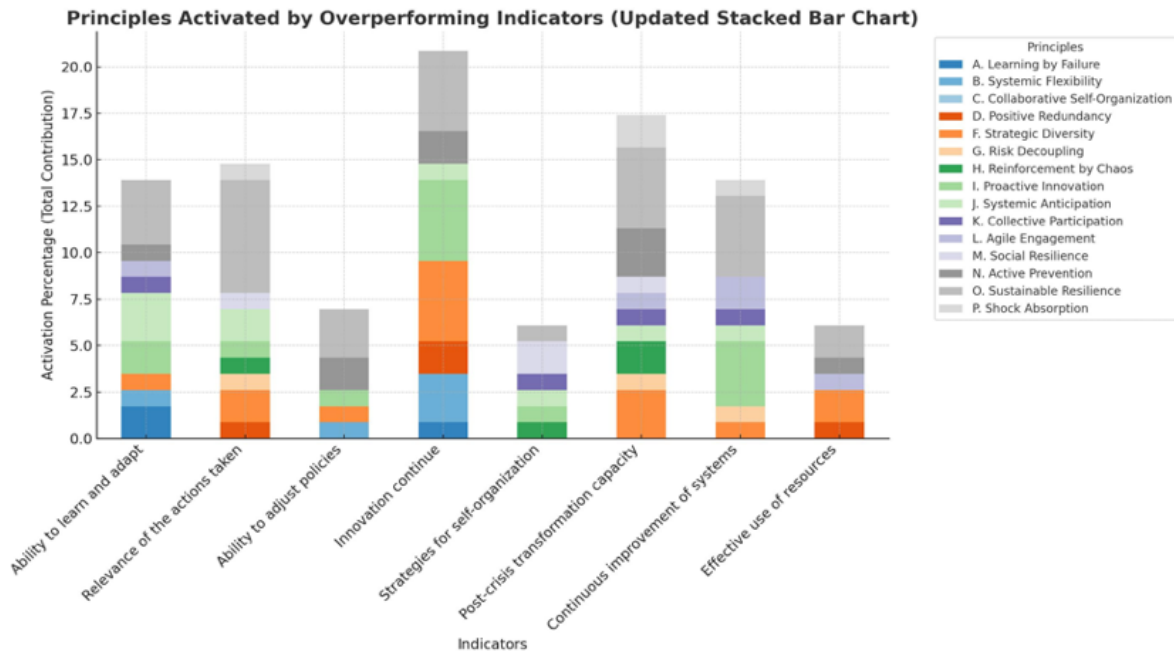

*Figure 15. Distribution of the principles activated by the outperforming indicators. Source: Author*

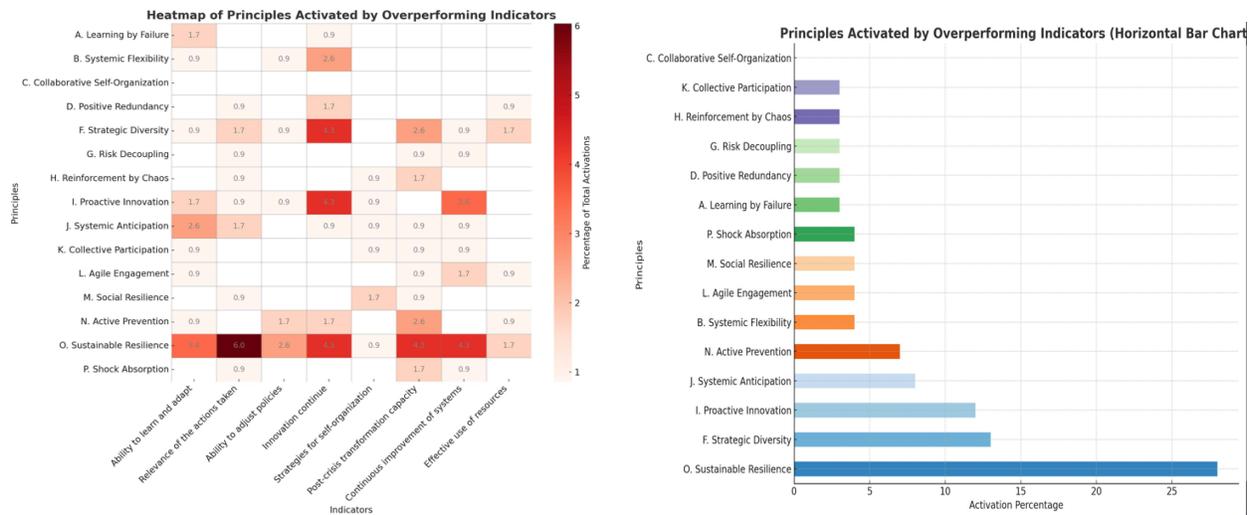

*Figure 16. Distribution of the principles activated by the outperforming indicators. Source: Author*

**Post-crisis transformation and strategic levers**. Activated in **Fukushima** and **New Orleans**, the indicator **Post-Crisis Transformation Capacity** represents 17% of total principle activations. It mobilizes five complementary principles: **O (Sustainable Resilience),** anchoring systemic optimization and circularity; **N (Active Prevention),** focused on proactive risk reduction; **F (Strategic Diversity)**, enabling multiple adaptive pathways; **P (Shock Absorption),** cushioning disruption and stabilizing critical functions; and **H (Reinforcement through Chaos),** leveraging controlled disruption to enable structural renewal. Together, these principles reflect a comprehensive approach to converting shocks into durable gains.

Across the full set of outperforming indicators, three principles stand out—**O, F, and I**—accounting for **over 55%** of all activations. **Principle O** is the only one activated across **all eight** outperforming indicators, confirming its systemic centrality in antifragile trajectories, even where it is not the dominant driver.

**Principle F**, while sometimes operating in the background, proves decisive: contributing **20% to Continuous Innovation** and **15% to Post-Crisis Transformation**, it preserves optionality and flexibility under uncertainty.

**Principle I (Proactive Innovation),** responsible for 12% of total activations, provides the forward-looking engine for integrating adaptive solutions, especially in indicators linked to innovation and system efficiency.

This layered reading of principle–indicator relationships clarifies the distinct roles—and complementarities—of **O, F, N, and I**. Together, they form the structural "grammar" of antifragility. The next step is to decode how these principles **co-activate and interact**, since effective strategies emerge not from isolated mechanisms, but from their orchestration.

*Analysis of the Co-Activated Principles of Outperforming Indicators*

The co-activation analysis, based on **57 best practices** associated with outperforming indicators, reveals the systemic architecture of urban antifragility. The heatmap (Figure 17) highlights the most recurrent binomials and confirms the strategic centrality of four principles: **F (Strategic Diversity), O (Sustainable Resilience), I (Proactive Innovation), and N (Active Prevention)**. Together, these principles appear in **47 of the 57** co-activated pairs (**82% of cases**), underscoring their foundational role. **O** alone is present in **32 pairs (56%)**, confirming its dual function as both a conceptual anchor and an operational driver of antifragile transition. Overall, the results show that antifragility arises not from isolated measures, but from **orchestrated complementarities** among core principles.

Figure 18 provides a synthetic view of how antifragility principles are **co-activated** within the best practices associated with **outperforming indicators**. Each principle is represented on the outer ring, where the width of the segment and the number in parentheses indicate how frequently that principle appears in co-activation events. The connecting ribbons (chords) show which principles are activated together within the same best practice; ribbon thickness reflects the recurrence of each pairing. The diagram reveals a clearly structured architecture: a limited set of principles concentrates most co-activations, with **Sustainable Resilience (O / O')** emerging as the main hub, repeatedly coupled with other high-leverage principles such as **Strategic Diversity (F), I (Proactive Innovation)** and **Active Prevention (N)**. Overall, the figure confirms that the strongest antifragile performances are not produced by isolated principles, but by **recurrent complementarities**—stable combinations in which resilience-oriented measures are systematically reinforced by diversity, prevention, innovation, and participation mechanisms.

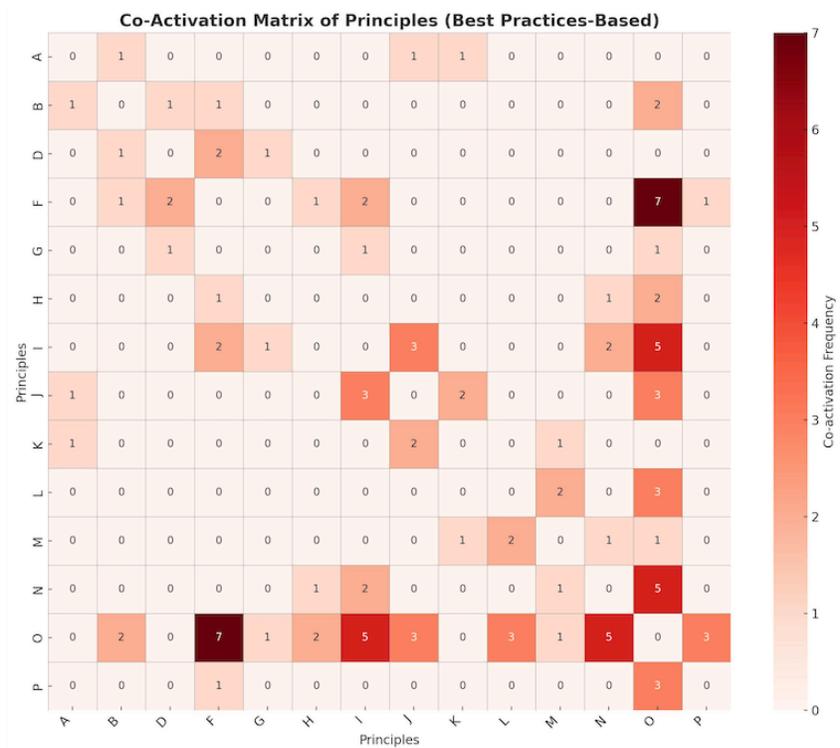

*Figure 17. Co-activation heatmap of the Principles activated by the Outperforming Indicators. Source: Author*

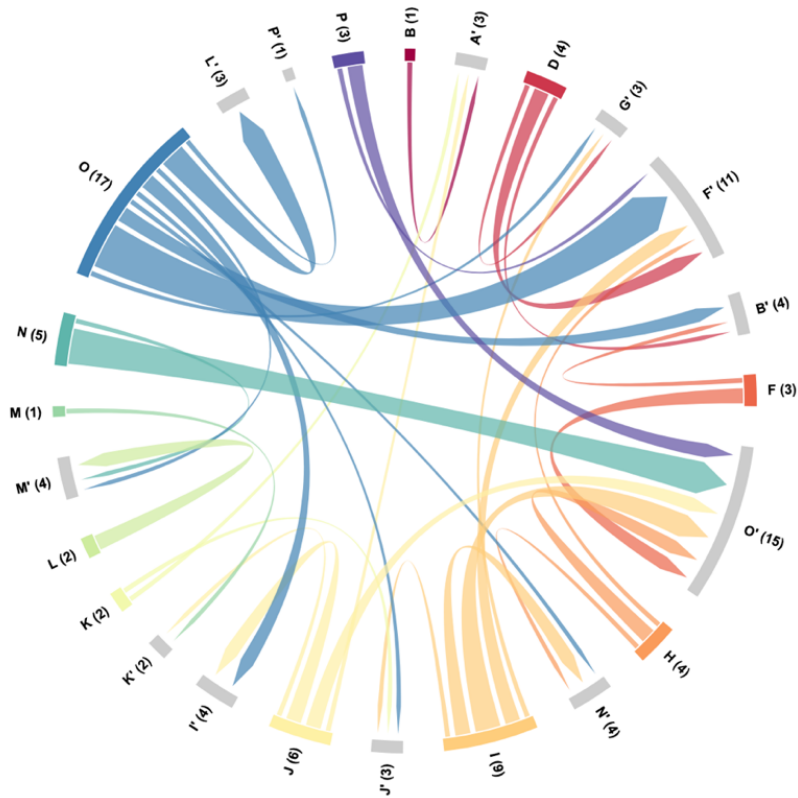

*Figure 18. Co-activation chord of the Principles activated by the Outperforming Indicators. Source: Author*

*From Clusters to Urban Strategies*

Figure 17 and Figure 18 reveal three coherent clusters of principles, identified through high-frequency co-occurrences (3 to 7), that structure the architecture of antifragile urban strategies. Figure 19 consolidates these results by visualizing the 15 antifragility principles and the three clusters that underpin the transformative dynamics observed in the best-performing cities.

First, a **"hard core"** of four principles forms the backbone of antifragility: **F (Strategic Diversity), O (Sustainable Resilience), I (Proactive Innovation), and N (Active Prevention)**. Together, they secure flexibility, long-term robustness, continuous evolution, and risk anticipation. Their strong interconnections—particularly **F–O (7), I–O (5), and N–O (5)**—confirm their central role in organizing adaptive and sustainable crisis responses.

Second, a set of **"reinforcement" principles—J (Systemic Anticipation), L (Agile Mobilization), and P (Shock Absorption)**—strengthens the core by improving responsiveness and shortening adaptation time.

Frequently co-activated with the backbone, notably through **J–O (3), L–O (3), P–O (3), and J–I (3)**, these principles operate as accelerators that enable faster and more effective implementation.

Third, a group of **"bridge" principles**—less recurrent but strategically significant—includes **B (Institutional Flexibility), D (Redundancy), G (Risk Decoupling), H (Reinforcement through Chaos), K (Collective Participation), and M (Social Resilience)**. When connected to the core, they allow strategies to be tailored to local conditions, strengthening territorial fit and reinforcing systemic coherence.

Taken together, these clusters show that antifragility emerges from a dynamic interplay between **foundational principles**, **operational enablers**, and **contextual bridges**—forming a framework that is both flexible and robust, and capable of supporting transformative urban resilience.

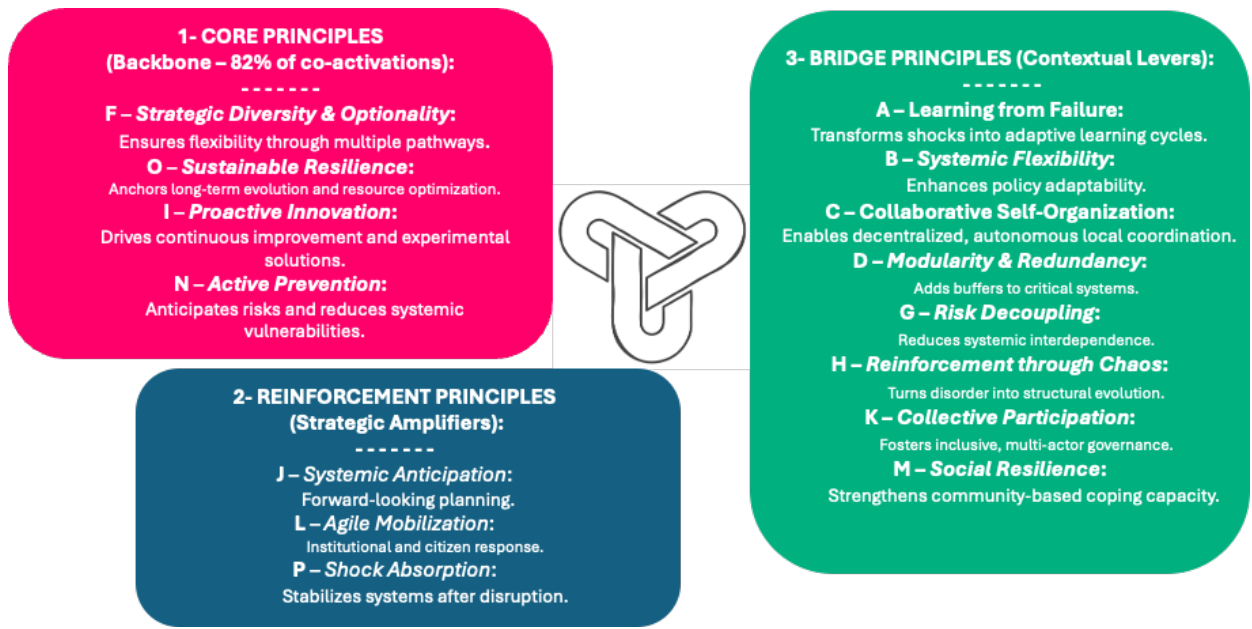

*Figure 19. 15 Antifragile Principles: Dynamics Driving Urban Transformation. Source: Author*

**Full activation of the theoretical framework.** With the exception of **C (Collaborative Self-Organization)**, all theoretical principles are activated in at least one co-occurrence pair, supporting the overall robustness of the antifragility model. Lower-frequency activations should not be interpreted as a lack of relevance; they may reflect limitations of the analytical tool, case-selection effects, or the fact that some principles are highly context-dependent. Even when less recurrent, these principles retain strategic value, often exerting influence through targeted contributions tailored to local needs and constraints.

**Strategic implications for urban planning.** The findings confirm that antifragility emerges from the **orchestration of co-activated principles**, rather than from isolated actions. Leading cities—**Singapore, Barcelona, Paris, and Fukushima**—consistently mobilize a core set (**F–O–I–N**), complemented by **J, L, and/or P** depending on priorities and shock profiles (e.g., **O and N** for climate-related risks; **F and I** to support economic recovery and innovation; **J and P** in post-disaster contexts). This cluster logic supports a **modular planning approach**: first consolidating the resilience base (**O, F**), then integrating innovation (**I**), anticipation (**J**), and context-specific levers (such as **N** and **K**) to strengthen legitimacy and implementation capacity. Antifragile strategies therefore take shape through dynamic combinations in which **O, I,** and **N** act as primary drivers, reinforced by **F, L, J, and P**, to build adaptive, scalable, and transformative urban systems—thereby refining the framework and strengthening its operational relevance.

**Discussion**

The analysis unfolded through successive lenses, progressively bridging the gap between theoretical models and observed urban transformations. Moving from dominant strategies to activated principles, each step sharpened the understanding of antifragile dynamics, using a funnel-like approach to reveal the mechanisms and combinations most consistently associated with high performance.

**Key findings and Strategic Dynamics.** Across the 26 cities assessed, **resilience enhanced by innovation and technology** emerged as the most effective strategy (**86.9/100**), adopted by seven cities—five of which rank in the top ten. This pattern, confirmed at each analytical layer, shows that antifragile trajectories depend on **interdimensional alignment**, particularly across **learning, innovation, and post-crisis transformation**.

Cities classified as antifragile consistently perform strongly on these dimensions, demonstrating an ability to adapt continuously and to convert disruption into a driver of strategic renewal.

The focus on Best Practices further highlighted a core set of four principles—**O (Sustainable Resilience), F (Strategic Diversity), I (Proactive Innovation), and N (Active Prevention)**. Recurrent and mutually reinforcing, these principles form the backbone of urban antifragility. Around this core, secondary principles operate as adaptive supports, activated in response to contextual needs and contributing to overall systemic robustness.

**Dominant, Supporting and Latent Principles.** The results confirm **O, F, I, and N** as the structural pillars of antifragile strategies. A second tier of supporting principles—**J (Systemic Anticipation), L (Agile Mobilization), and P (Shock Absorption)**—strengthens responsiveness and operational agility, even if they are not consistently catalytic on their own. **P**, in particular, becomes more prominent in high-performing configurations, while **K (Collective Participation)**—initially salient—appears less frequently in the dominant pairings.

Figure 17 and Figure 18 illustrate the prominence of these principles in co-activation patterns, reinforcing their systemic relevance. By contrast, several principles frequently emphasized in the literature—**C (Collaborative Self-Organization), H (Reinforcement through Chaos), D (Redundancy), G (Risk Decoupling), and A (Learning through Failure)**—remain marginal or absent in the dominant configurations (Figure **20**). Their limited activation does not invalidate their potential contribution; rather, it may reflect weaker formalization in city strategies, context-specific deployment that is not captured by the indicators, or measurement and classification constraints.

As illustrated in Figure 20, the gap between theoretical prominence and empirical activation—particularly for **C** and **H**—suggests the need to further examine **informal governance**, **grassroots adaptation**, and other latent dynamics that may not be captured by conventional indicators. Conversely, **N** and **P**, sometimes treated as secondary, show strong activation in crisis management, highlighting their role in limiting structural impacts and creating the conditions for transition. Overall, these results call for a recalibration of certain theoretical assumptions and underscore the value of **empirical validation** when operationalizing antifragile urban strategies.

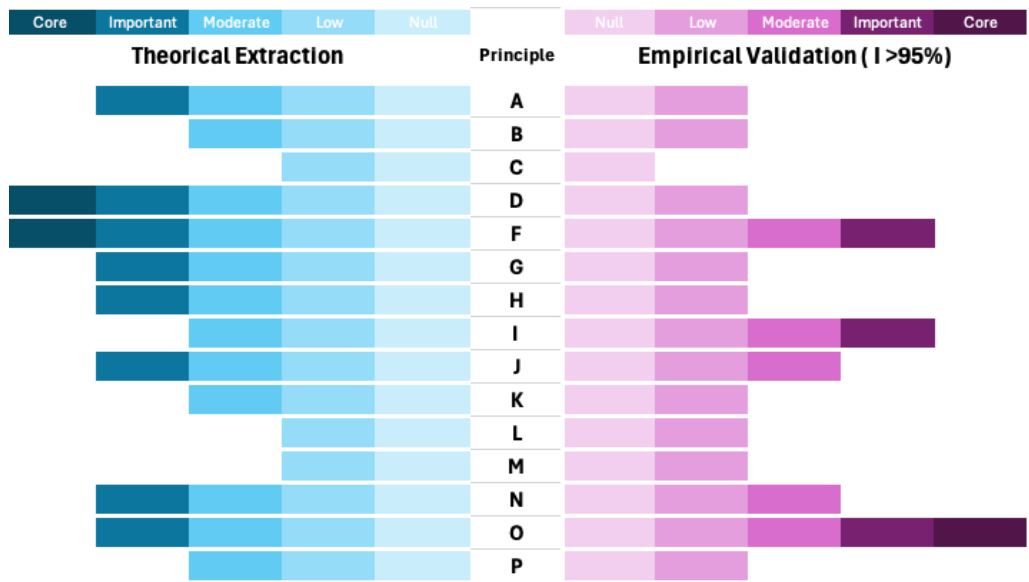

*Figure 20. 15 Principles: Theoretical vs. Empirical – Toolbox Validation. Source: Author*

**Detection Deficit vs Contextual Relevance.** The divergence between theoretical prominence and empirical visibility—particularly for principles such as **C (Self-Organization)**, **H (Reinforcement through Chaos)**, and **G (Risk Decoupling)**—can be explained, in part, by a **detection deficit**. These principles lack standardized, widely available indicators and are therefore difficult to capture through conventional urban data systems. Their conceptual foundations, often interdisciplinary and experimental, are not easily translated into mainstream policy instruments or benchmarking tools.

Context is also decisive. Some principles are activated primarily under exceptional conditions—severe shocks, cascading failures, or long-term recovery phases—and may therefore remain underrepresented when such episodes are limited or absent in the sample. **Temporality** further complicates interpretation: certain concepts, though influential in theory (e.g., **H**), have not yet matured operationally, while others may be implemented in practice without being explicitly codified in academic or policy language. Conversely, emerging approaches—such as data-driven prevention—may be insufficiently reflected in older literature, creating a lag in theoretical recognition. Meanwhile, the salience of established principles may shift as urban risk profiles and governance challenges evolve. Together, these tensions highlight the need for **adaptive frameworks** capable of capturing both emerging practices and latent dynamics.

**The power of pairs and co-activations.** Antifragility does not arise from isolated principles but from **synergistic combinations**. The recurring pairs **F–O**, **I–O**, and **N–O** reveal a strategic architecture in which diversity, innovation, sustainable resilience, and anticipation reinforce one another. Urban robustness, in this sense, is produced through orchestrated interactions rather than the dominance of a single "star" principle.

Co-activations reflect **adaptive intelligence**: some cities pair **L (Agile Mobilization)** with **I (Proactive Innovation)**, while others couple **N (Active Prevention)** with **O (Sustainable Resilience)**—each configuration calibrated to local constraints, capacities, and shock profiles. This flexibility confirms that antifragility resists standardization; it is continuously recomposed through strategic pairings that both reduce vulnerabilities and open pathways for evolution. These findings argue for a framework that prioritizes **relational logic** over static lists. Identifying high-performing pairs provides cities with an actionable roadmap to allocate resources, align actors, and synchronize transformations. In this perspective, antifragility becomes operational: an acceleration mechanism forged through strategic complementarities.

**Structuring the conceptual Framework.** The convergence of theoretical and empirical insights calls for a framework that balances conceptual richness with clear priorities. The core principles—**F, I, O,** and **N**—form the foundation. Less frequently activated principles—**C, H, A** and **G,** —retain strategic potential and are better positioned as **contextual levers**, particularly relevant in extreme situations, experimental policy environments, or innovation-driven transitions. Several orientations follow from the analysis:

- **Preserve plurality:** rather than excluding under-activated principles, integrate them as "satellite" elements whose activation is conditioned by context.
- **Clarify transversal roles:** principles such as **B (Systemic Flexibility)**, even when less visible, may operate as cross-cutting enablers across multiple strategies.
- **Refine measurement:** develop tailored indicators for complex dynamics (e.g., chaos, self-organization) through fieldwork, process tracing, and network analysis.
- **Align with SDGs:** link each principle to relevant global targets (e.g., **SDG 9**, **SDG 11**) to strengthen institutional uptake and comparability.

Ultimately, the robustness of the framework lies in its capacity to **co-evolve**. New principles will emerge, and existing ones will gain analytical depth through improved instrumentation. Anchored in sustained dialogue between theory and practice, urban antifragility is not a static model, but a living architecture of transformation.

**Conclusions**

This study advances urban antifragility from a conceptual proposition to an evidence-based operational framework. Starting from fifteen theoretical principles, we empirically assessed 26 cities and positioned their trajectories along a fragility–robustness–resilience–antifragility continuum using a two-stage diagnostic process: (i) benchmarking crisis response through the SRS index and (ii) validating longer-term transformation through the UDT framework.

Crucially, antifragility is treated as a condition of **balanced systemic performance** rather than an aggregated score alone, requiring each UDT dimension to exceed a defined threshold. This rule-based validation strengthens the interpretability of results and clarifies the difference between cities that recover effectively and those that use disruption to produce structural gains.

Three substantive findings stand out. First, the comparative strategy assessment shows that **innovation-driven crisis response** is consistently associated with higher effectiveness: "resilience enhanced by innovation and technology" emerges as the strongest typology (**86.9/100**), adopted by seven cities, five of which appear among the top SRS performers.

Second, the UDT diagnosis confirms **six antifragile trajectories**, including Singapore, Paris, Barcelona, Tokyo, Fukushima, and New Orleans, while other high-performing SRS cases remain classified as resilient due to weaker balance across dimensions.

Third, the translation from outcomes to mechanisms—through best-practice mapping and co-activation analysis—reveals that antifragile performance is driven less by isolated interventions than by **recurrent complementarities** among principles.

Mechanistically, the analysis identifies a coherent strategic architecture organized around a "hard core" of four principles: **O (Sustainable Resilience), F (Strategic Diversity), I (Proactive Innovation), and N (Active Prevention)**.

These Principles dominate co-activations and structure most high-performing pairings, indicating that antifragility emerges through combined capacities to optimize systems (O), preserve optionality (F), innovate continuously (I), and reduce risk proactively (N). A second tier of "reinforcement" principles—anticipation, agile mobilization, and shock absorption—acts as accelerators that shorten response times and support implementation under stress. Additional "bridge" principles provide contextual tailoring, helping cities adapt the core architecture to distinct hazard profiles, governance configurations, and socio-economic conditions.

Beyond validating the framework, the study yields an actionable planning implication: cities can operationalize antifragility through a **modular strategy logic**—consolidating a resilience base (O, F), then integrating innovation and anticipation capacities (I, J/L), and finally activating context-specific levers (e.g., prevention, participation) to ensure legitimacy and implementation capacity.

In this sense, **antifragility becomes a structured, adaptable toolbox grounded in demonstrated performance, and a foundation for an SDG-aligned application model that links principles, high-impact pairings, and measurable indicators.**